\documentclass[twocolumn,english]{revtex4-1}
\usepackage[T1]{fontenc}
\usepackage[latin9]{inputenc}
\usepackage{geometry}
\usepackage{babel}
\usepackage{float}
\usepackage{amsmath}
\usepackage{graphicx}

\makeatletter

\floatstyle{ruled}
\newfloat{algorithm}{tbp}{loa}
\providecommand{\algorithmname}{Algorithm}
\floatname{algorithm}{\protect\algorithmname}

\usepackage{algorithm,algpseudocode}

\makeatother

\begin{document}
\title{Information-geometric optimization with natural selection}
\author{Jakub Otwinowski}
\affiliation{Max Planck Institute for Dynamics and Self-Organization}
\email{jakub.otwinowski@ds.mpg.de}
\author{Colin H. LaMont}
\affiliation{Max Planck Institute for Dynamics and Self-Organization}

\begin{abstract}
Evolutionary algorithms, inspired by natural evolution, aim to optimize
difficult objective functions without computing derivatives. Here
we detail the relationship between population genetics and evolutionary
optimization and formulate a new evolutionary algorithm. Optimization
of a continuous objective function is analogous to searching for high
fitness phenotypes on a fitness landscape. We summarize how natural
selection moves a population along the non-euclidean gradient that
is induced by the population on the fitness landscape (the natural
gradient). Under normal approximations common in quantitative genetics,
we show how selection is related to Newton's method in optimization.
We find that intermediate selection is most informative of the fitness
landscape. We describe the generation of new phenotypes and introduce
an operator that recombines the whole population to generate variants
that preserve normal statistics. Finally, we introduce a proof-of-principle
algorithm that combines natural selection, our recombination operator,
and an adaptive method to increase selection. Our algorithm is similar
to covariance matrix adaptation and natural evolutionary strategies
in optimization, and has similar performance. The algorithm is extremely
simple in implementation with no matrix inversion or factorization,
does not require storing a covariance matrix, and may form the basis
of more general model-based optimization algorithms with natural gradient
updates. 
\end{abstract}
\maketitle

\section*{Introduction}

Finding the optimal parameters of a high dimensional function is a
common problem in many fields. We seek protein conformations with
minimal free energy in biophysics, the genotypes with maximal fitness
in evolution, the parameters of maximum likelihood in statistical
inference, and optimal design parameters in countless engineering
problems. Often derivatives of the objective function are not available
or are too costly, and derivative-free algorithms must be applied. 

Evolutionary optimization algorithms (EA) use a population of candidate
solutions to generate new candidate solutions for the objective ``fitness''
function. In particular, genetic algorithms (GA) and evolution strategies
(ES) are two classes of EAs most directly inspired by the Wright-Fisher
and Moran models from population genetics \cite{simon_evolutionary_2013,gillespie_population_2004}.
GAs are initialized with some population of genotypes, representing
candidate solutions, and use some form of stochastic reproduction,
incorporating a bias known as selection. Among the different selection
schemes, fitness-proportionate selection is equivalent to natural
selection in population genetics.

Stochasticity of reproduction may be helpful in overcoming local optima,
and noise in fitness. In population genetics, stochasticity of reproduction
is known as genetic drift, and has the important effect of scaling
the strength of selection inversely with the magnitude of stochasticity
\cite{gillespie_population_2004}. Stochasticity also causes the loss
of many genotypes, even if they have high fitness. For example in
the strong selection weak mutation regime of the Moran model, the
probability that a single genotype will sweep a population (fixation)
is proportional to its selective advantage, and there is a 99\% chance
a genotype with a one percent fitness advantage will go extinct \cite{gillespie_population_2004}. 

Some form of deterministic selection in optimization is desirable
to not waste computational effort, and without stochasticity a population
based algorithm will still be robust to noise in fitness since a population
effectively integrates information over some region of the fitness
landscape. Some ES and GAs use deterministic rank based selection
which removes individuals below some threshold. However, such truncation
selection is very coarse, and does not affect proportionally the genotypes
that survive. 

Many population based algorithms, including ES and estimation of distribution
algorithms (EDA), are based on drawing a population of candidate solutions
from a parameterized distribution $\mathcal{P}(\theta)$ and iteratively
updating the parameters $\theta$ \cite{hauschild_introduction_2011}.
The basic approach is to move the parameters in the direction of the
gradient of the mean fitness: $\nabla_{\theta}F$. To account for
the uncertainty of the parameters many algorithms move the parameters
in the direction of the natural gradient \cite{amari_why_1998}, $g^{-1}\nabla_{\theta}F$,
where $g^{-1}$ is the inverse of the fisher information, which can
be estimated from the population. The popular covariance matrix adaptation
ES algorithm (CMA-ES) \cite{hansen_adapting_1996,hansen_cma_2006}
and related natural evolution strategies (NES) \cite{wierstra_natural_2008,sun_efficient_2009,glasmachers_exponential_2010,wierstra_natural_2014}
parameterize a population as a normal distribution, and use samples
from the distribution to update the mean and covariance with a natural
gradient descent step. More generally, natural gradients describe
ascent of the fitness landscape in terms of information geometry,
and their use characterizes a wide class of information-geometric
algorithms \cite{ollivier_information-geometric_2011}. These algorithms
differ from GAs and population genetics, in that there are no selection
or mutation operators applied directly to individuals in a population.
\begin{figure*}[t]
\begin{centering}
\includegraphics[width=0.6\textwidth]{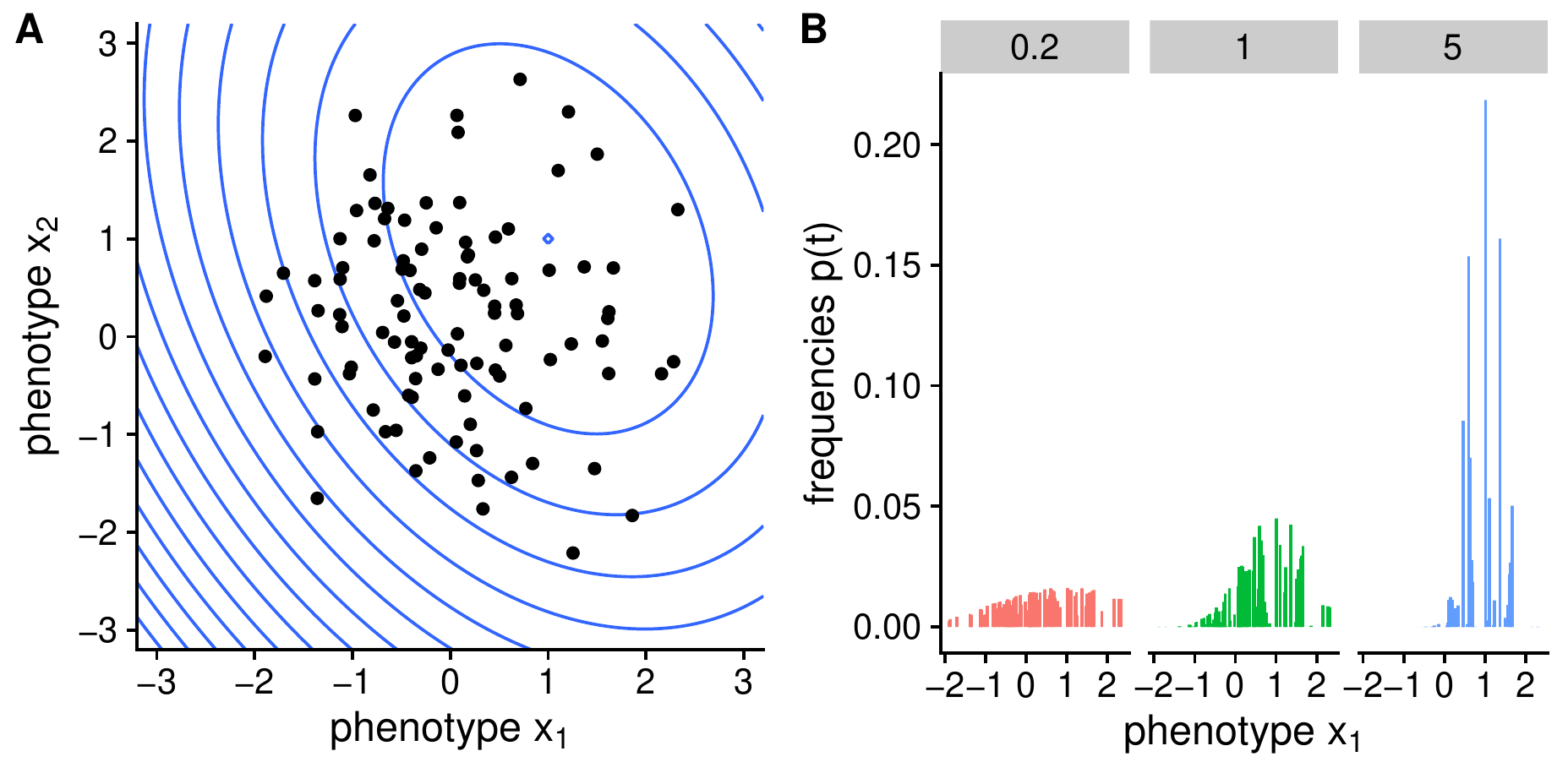}
\par\end{centering}
\caption{\label{fig:example}A) An example of 100 variants in a 2D phenotype
space on a quadratic fitness landscape (blue contours). B) Frequencies
evolve over time according to eq.~\ref{eq:repl2}, with $t=0.2$
(red), $t=1$ (green) and $t=5$ (blue). }
\end{figure*}

Here, we point out that the natural gradient used in information-geometric
optimization also appears in natural selection. Under normally distributed
phenotypes, as is done in multivariate quantitative genetics, we show
how selection is related to Newton's method in optimization. Then,
we describe how intermediate levels of selection are best for optimization,
and how mutation and recombination generate new variants without having
to explicitly sample from the distribution. Finally, we develop a
proof of principle quantitative genetic algorithm (QGA) which combines
selection, recombination, and a form of adaptive selection tuning
that shrinks the population towards an optimum. In contrast to GAs,
QGA has deterministic selection, and compared to CMA-ES and NES, it
is much simpler and does not store a covariance matrix.

\section*{Natural selection gradient}

We begin by considering a population of infinite size, but with a
finite number of unique phenotypes. Each unique variant $i$ has a
continuous multivariate phenotype $\boldsymbol{x}_{i}$ (a vector),
with frequency $p_{i}$ and growth rate, or fitness $f(\boldsymbol{x}_{i})$,
independent of time and frequencies. In the context of optimization,
phenotypes are candidate solutions, and fitness is the objective function
to be maximized. Classical replicator dynamics, leaving out mutation
and genetic drift, describe the change in frequencies as
\begin{equation}
\frac{dp_{i}}{dt}=p_{i}\left(f(\boldsymbol{x}_{i})-F\right),\label{eq:repl1}
\end{equation}
with mean fitness $F=\sum_{i}p_{i}f(\boldsymbol{x}_{i})$. In stochastic
descriptions, these dynamics describe the expected changes due to
selection. 

In the absence of other processes, frequencies can be integrated over
time resulting in
\begin{equation}
p_{i}(t)=p_{i}(0)\frac{1}{Z_{t}}e^{tf(\boldsymbol{x}_{i})},\label{eq:repl2}
\end{equation}
with normalization $Z_{t}$ ensuring the probabilities sum to one.
At long times, the phenotype distribution will concentrate on high
fitness phenotypes until the highest fitness phenotype reaches a frequency
of unity. The change in mean fitness equals the fitness variance (Fisher's
theorem), and higher moments evolve as well. As an example we show
100 variants in a quadratic fitness landscape and how their frequencies
change over time (Fig.~\ref{fig:example}). 

Remarkably, replicator dynamics can be rewritten in terms of information
geometry \cite{amari_information_2016}. Frequencies can be considered
as the parameters of a discrete categorical distribution, and selection
moves them in the direction of the covariant derivative \cite{harper_information_2009,mustonen_fitness_2010},
(also known as the natural gradient \cite{amari_why_1998}), 
\[
\frac{d\boldsymbol{p}}{dt}=\boldsymbol{g}^{-1}\nabla_{\boldsymbol{p}}F,
\]
where $\boldsymbol{p}$ is the vector of (linearly independent) frequencies,
$\nabla_{\boldsymbol{p}}F$ is the vector of partial derivatives,
and $\boldsymbol{g}^{-1}$ is the inverse of the fisher information
metric of the categorical distribution, which defines distances on
the curved manifold of probability distributions (see appendix \ref{sec:Selection}).
Selection changes the frequencies in the direction of steepest ascent
in non-euclidean coordinates defined by the geometry of the manifold
of the distribution.

The natural gradient is independent of parameterization, and therefore,
if the distribution over $\boldsymbol{x}$ can be approximated by
another distribution, selection will change those parameters in the
direction of their natural gradient. This can be demonstrated by projecting
onto a normal phenotype distribution, as is assumed in classic multivariate
quantitative genetics. The population mean $\boldsymbol{\mu}$ and
population covariance matrix $\boldsymbol{\Sigma}$ parameterize the
distribution, and selection changes the mean as (\cite{lande_quantitative_1979,nourmohammad_evolution_2013},
appendix \ref{sec:Selection})
\begin{equation}
\frac{d\boldsymbol{\mu}}{dt}=\boldsymbol{\Sigma}\nabla_{\boldsymbol{\mu}}F,
\end{equation}
where $\boldsymbol{\Sigma^{-1}}$ is the associated Fisher information
metric. Similarly, the covariance follows a natural gradient with
a more complex metric (appendix \ref{sec:Selection}). If phenotype
covariance reaches zero, then the population is monomorphic and there
is no selection. However, an alternative population genetics model
in the strong selection weak mutation regime can search a fitness
landscape with limited diversity, with the mutation covariance matrix
serving as a metric (Appendix \ref{sec:SSWM}).

For a finite amount of time, the frequencies have Boltzmann form and
the parameters trace a path on the manifold of distributions following
the natural gradient. Exponential weights lead to natural gradient
updates and are found in many optimization algorithms beyond GAs,
such as simulated and population annealing \cite{kirkpatrick_optimization_1983,amey_analysis_2018}.
In contrast, CMA-ES/NES use a population to update the parameters
of a normal distribution in a natural gradient step, and do not track
frequencies. 

\section*{Selection as a second order update}

To underscore how natural gradients are inherently second order optimization
methods, we show how natural selection is related to a second order
gradient update when phenotypes are normally distributed. The normal
approximation is valid for quadratic fitness landscapes, or when the
taylor expansion of $f(\boldsymbol{x})$ around the mean phenotype
$\boldsymbol{\mu}$ to second order is a good approximation.  The
mean and covariance change as (appendix \ref{sec:newton})
\begin{align}
\boldsymbol{\mu}_{t}-\boldsymbol{\mu}_{0} & =t\boldsymbol{\Sigma}_{0}\nabla f\nonumber \\
\boldsymbol{\Sigma}_{t}^{-1}-\boldsymbol{\Sigma}_{0}^{-1} & =t\boldsymbol{C}
\end{align}
where subscripts indicate dependence on time, $\nabla f$ are the
partial derivatives of $f(\boldsymbol{x})$ evaluated at $\boldsymbol{\mu}_{0}$,
$\boldsymbol{C}$ is the curvature, that is the negative of the matrix
of second derivatives, evaluated at $\boldsymbol{\mu}_{0}$. The change
in mean is a finite natural gradient step, while the covariance aligns
itself with the curvature with increasing time. Combining the two
equations yields

\begin{equation}
\boldsymbol{\mu}_{t}-\boldsymbol{\mu}_{0}=\left(\frac{1}{t}\boldsymbol{\Sigma}_{0}^{-1}+\boldsymbol{C}\right)^{-1}\nabla f.\label{eq:newt-1}
\end{equation}
 For large $t$, $\boldsymbol{\mu}_{t}-\boldsymbol{\mu}_{0}\to\boldsymbol{C}{}^{-1}\nabla f$
corresponding to an iteration of Newton's method, as long as the normal
approximation holds (see below). For small and intermediate $t$,
selection functions as a form of regularized Newton's method, where
$t$ determines how much to weigh the prior, that is the initial distribution.
Since $\boldsymbol{\Sigma}_{t}$ is always positive-definite, the
population cannot converge to a saddle point in fitness, which is
possible in Newton's method.

\section*{Intermediate selection is most informative}

\begin{figure*}[t]
\centering{}\includegraphics[width=1\textwidth]{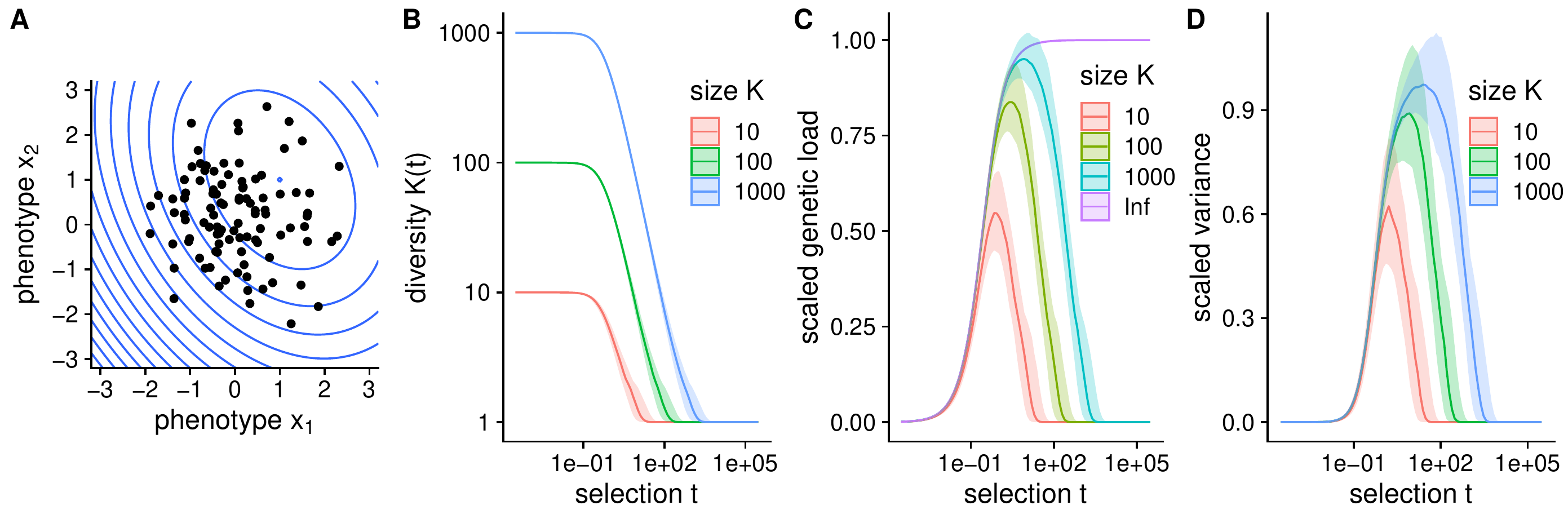}\caption{\label{fig:finiteselection}Selection reduces diversity and intermediate
levels of selection are most informative of the landscape. Fitness
landscape is show in Fig.~\ref{fig:example}A. A) The effective number
of variants, $K_{t}$, shrinks with increased selection and depends
on the total number of variants (colors). Populations are drawn 1000
times from a standard normal with solid lines indicating the median
value and shaded regions indicating the 50\% confidence interval.
B) Scaled genetic load $t(F^{\dagger}-F)$, that is the gap between
the maximum observed fitness $F^{\dagger}$ and mean fitness $F$,
has a peak value at intermediate selection due to the loss of diversity
at strong selection. Purple line indicates continuous limit of a normal
distribution (Appendix \ref{sec:Genetic-load}). C) Fitness variance
(multiplied by $t^{2}$ to make it unit-less) also peaks at some intermediate
level of selection.}
\end{figure*}
 Natural selection moves the distribution along a manifold shaped
by the fitness landscape. However, selection does not introduce any
new phenotypes, and reduces diversity by biasing frequencies towards
high fitness phenotypes. Diversity can be quantified by population
entropy $S_{t}=-\sum_{i}p_{i}(t)\log p_{i}(t)$, which summarizes
the distribution of frequencies. The exponential of entropy $K_{t}=e^{S_{t}}$
defines an effective number of variants, such that $1\le K_{t}\le K_{0}$,
and $K_{0}$ is the number unique variants under uniform initial conditions
$p_{i}(0)=1/K_{0}$. Diversity shrinks rapidly as selection increases,
depending on the starting point $K_{0}$ (Fig.~\ref{fig:finiteselection}B). 

There is an inherent tradeoff in choosing the strength of selection
$t$, in that small $t$ weakly biases the frequencies towards high
fitness, and large $t$ has a low effective number of variants. If
the variants are regarded as samples drawn from an underlying distribution,
fewer samples leads to higher error in estimating parameters, such
as mean and covariance.

For large $t$ and small $K_{t}$ the normal approximation breaks
down, and eqn.~\ref{eq:newt-1} does not hold even if fitness is
purely quadratic. The breakdown of normality is reflected in the moments
of the fitness distribution. The gap between the mean and maximum
fitness $F^{*}-F$, known as genetic load, shrinks as selection increases,
but under normal phenotype distributions, $F^{*}-F\to\frac{D}{2t}$
where $D$ is the dimensionality of phenotypes (Appendix \ref{sec:Genetic-load}).
Fitness has units of $t^{-1}$, as evident from eqns.~\ref{eq:repl1}
and \ref{eq:repl2}. The unit-less approximate scaled load $t(F^{\dagger}-F)$
where $F^{\dagger}$ is the maximum observed fitness, is zero when
mean fitness approaches the maximum, and reaches a peak around some
intermediate level of selection (Fig \ref{fig:finiteselection}C).
Similarly, the fitness variance scaled by $t^{2}$ has a peak at intermediate
selection (Fig \ref{fig:finiteselection}D) since fitness variance
must be zero at high selection. Some intermediate level of selection
is most informative of the curvature of the fitness landscape, in
that it balances biasing towards high fitness phenotypes and the effective
number of unique variants in the distribution.

\section*{Recombination efficiently generates diversity}

While selection moves the mean phenotype toward an optimum at the
cost of reducing diversity, diversity can be restored by mutation
and recombination processes that add new variants to the population.
Biologically, mutations and recombination are changes in genotypes
and their effect on phenotypes depends on the genotype-phenotype map.
Mutations generate new variants that can rise to some frequency after
selection if beneficial, and the expected cost or benefit of a mutation
depends on the fitness landscape. 

As an alternative to specifying a genotype-phenotype map, we define
a mutation as a small stochastic normally distributed change in phenotype
space. Intuitively, if the mutational distribution is similar to the
curvature of the fitness landscape then the cost should be minimal.
However, there is no reason to expect such an alignment, and if a
population is on a high dimensional ridge, then a mutation is very
likely to fall off the ridge and be very costly. For normally distributed
populations and mutations, the expected fitness cost is proportional
to the effective number of directions which are deleterious (Appendix
\ref{sec:Mutations}). 

From an optimization perspective, mutations are very inefficient in
that they rarely generate good solutions by being poorly aligned with
the fitness landscape. However, recombination has the remarkable property
of being adaptive to the fitness landscape. In population genetics
models with genotypes, recombination is known to quickly break down
correlations between sites in a genome until linkage equilibrium is
reached. Under these conditions the phenotype covariance and expected
fitness of recombined offspring is the same as the parent population
(\cite{lande_genetic_1980,neher_statistical_2011}). If the population
is on a high dimensional ridge and the population covariance is aligned,
recombined solutions would also be aligned with the ridge. However,
linkage equilibrium can only be reached on fitness landscapes without
interactions between genetic sites, which excludes most fitness landscapes,
including quadratic fitness-phenotype maps. 

For the purposes of optimization, we are free to define any recombination
operator. GAs typically use recombination in the form of a crossover
operator, similar to genetic recombination. However, a naive application
of crossover to phenotypes would destroy any covariance in the population.
Here we define recombination of phenotypes that preserves covariance.
A pair of distinct phenotypes, chosen by weighted random sampling
with weights $p_{1}(t)$ and $p_{2}(t)$ can be recombined as 
\begin{equation}
\boldsymbol{x}^{\prime}=\boldsymbol{\mu}_{t}+\frac{1}{\sqrt{2}}\left(\boldsymbol{x}_{1}-\boldsymbol{x}_{2}\right),
\end{equation}
which has the same expectation $\boldsymbol{\mu}_{t}$ and covariance
$\boldsymbol{\Sigma}_{t}$ as the parent population. This recombination
operator preserves normal statistics, and is a way of sampling from
an implicit normal distribution without estimating its parameters.
This operator resembles the mutation operator in differential evolution
optimization, which also preserves normal statistics for certain parameter
values \cite{storn_differential_1997}. However, the error in the
mean and covariance can be large when diversity is low due to sampling
noise, and the number of unique recombinants can be too limited. We
improve the quality of recombination by generalizing to a stochastic
sum over the entire population

\begin{equation}
\boldsymbol{x}^{\prime}=\boldsymbol{\mu}_{t}+\sum_{i}\eta_{i}\sqrt{p_{i}(t)}(\boldsymbol{x}_{i}-\boldsymbol{\mu}_{t}),\label{eq:recomb}
\end{equation}
where $\eta_{i}$ is an instance of a standard normal random variable.
This operator also conserves normal statistics, tuned by strength
of selection $t$, and efficiently generates new variants without
storing a covariance matrix. In comparison, CMA-ES/NES must store
a covariance matrix, which may be challenging in large problems. 

\section*{Quantitative genetic algorithm}

\begin{figure*}
\includegraphics[width=1\textwidth]{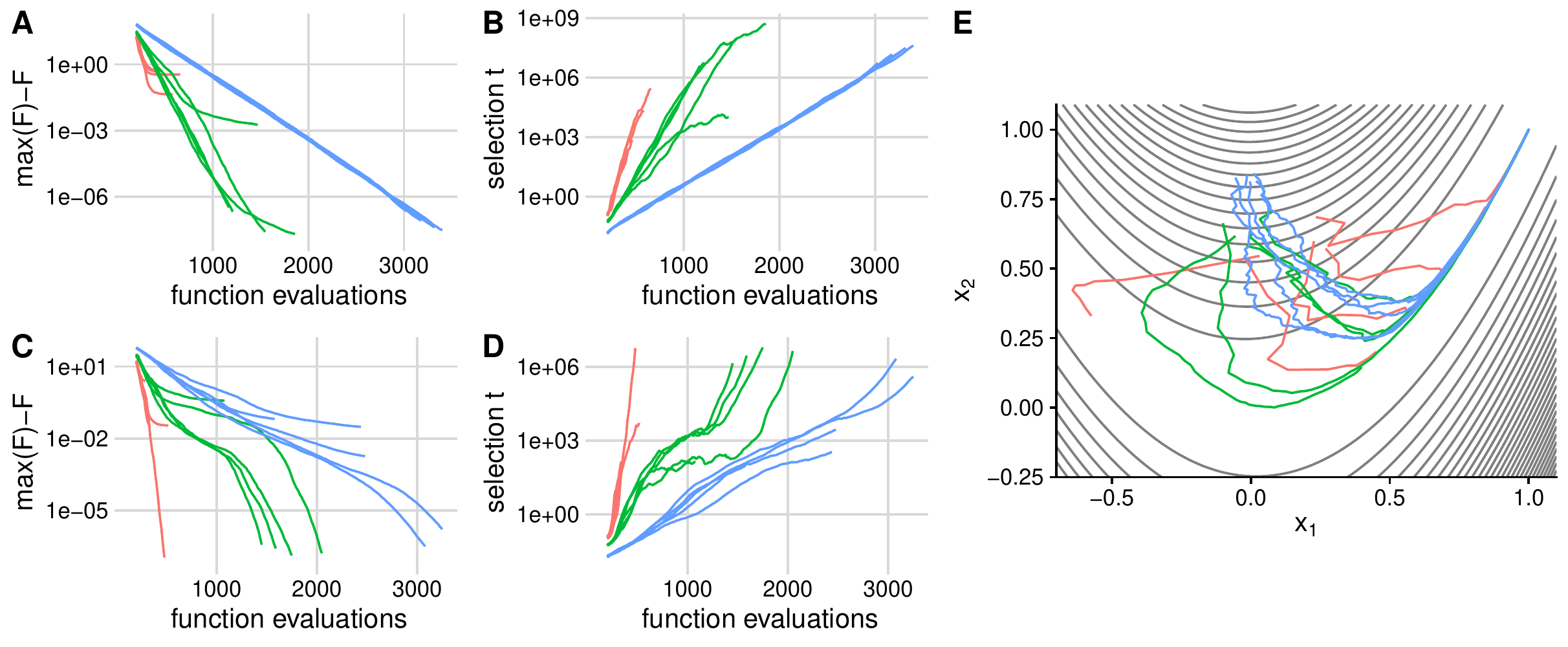}

\caption{\label{fig:opt2}QGA converges to the optimum when there is enough
diversity. Optimization on an ellipsoid (A,B) and a non-convex Rosenbrock
function (C,D,E), (see Methods) was carried out for three different
values of entropy $S$, 3 (green), 4 (red), and 5 (blue), and 5 independent
runs each. For the ellipsoid, distance to maximum fitness decreases
exponentially (A) and selection increases exponentially (B), unless
the population converges prematurely. For the Rosenbrock function,
the approach to the optimum fitness (C) and increase in selection
(D) are non-exponential as selection is adapted to keep the population
entropy $S$ at target values. E) Paths in phenotype space to the
optimum.}
\end{figure*}

\begin{algorithm}[H]
\begin{algorithmic}[1]
\State{choose hyperparameter $S$}
\State{draw population $\boldsymbol{x}_i$ for all $i$ from initial normal distribution}
\Repeat
\State{find $t$ such that $S=-\sum_{i}p_i(t)\log p_i(t)$} 
\State{sample standard normal variates $\eta_i$ for all $i$}
\State{add to population $\boldsymbol{x}^\prime \gets \sum_{i}\eta_i\sqrt{p_i(t)}(\boldsymbol{x}_i-\boldsymbol\mu_\infty)+\boldsymbol\mu_\infty$}
\Until{convergence criteria met}  
\end{algorithmic}
\caption{\label{alg:annealing}Quantitative genetic algorithm}
\end{algorithm}

We define a proof-of-principle \textit{quantitative genetic algorithm}
(QGA), as it is related to quantitative genetics, and is also a genetic
algorithm that is quantitative in the sense that it tracks frequencies
of types. QGA iterates between increasing selection to move the distribution
up the natural gradient, and phenotype recombination to generate variants.
To simplify matters, we generate only one variant per iteration, and
choose uniform initial conditions $p_{i}(0)=1/K_{0}$ for all variants,
even after a new variant is generated. This differs from population
genetics models where new variants are introduced at a small frequency.
In our algorithm, variants are never removed, but their frequencies
become small when they become irrelevant. 

A critical problem of EAs is how to increase, or sometimes decrease,
selection over the course of optimization to make sure it ends extremely
closely distributed around the optimum. One choice is a fixed schedule
of increases in selection, as in simulated annealing and population
annealing. However, choosing a good rate is challenging as increasing
selection too quickly will lead to premature convergence, and increasing
selection too slowly wastes resources. In addition, a constant rate
may not work well when different parts of the fitness landscape have
different curvatures. 

It is clear that some intermediate level of selection is needed, although
the peak in scaled genetic load or fitness variance is not necessarily
the right balance. We implement an adaptive strategy that keeps diversity
fixed and lets selection vary on its own. After a variant is generated,
population entropy will typically increase by a small amount since
entropy is an extensive quantity. Then we choose a new $t$ such that
the entropy is at a target value $S$, a hyper-parameter. This way
the population adapts to keep the diversity constant, with high fitness
variants driving selection higher as they are found. If $S$ is too
small the algorithm can still converge prematurely outside of the
local quadratic optimum. 

For generating variants, we use a modified recombination operator
which replaces $\boldsymbol{\mu}_{t}$ in Eq.~\ref{eq:recomb} with
the best variant seen thus far $\boldsymbol{\mu}_{\infty}$. Shifting
the mean of recombinants was found to perform much better in benchmarks
(see below and Fig.~S\ref{fig:benchmark1}). We provide an implementation
of QGA (Alg.~\ref{alg:annealing}) online, with minor implementation
details including a rudimentary test for premature convergence described
in Methods.

\section*{QGA performance }

 We demonstrate QGA on two simple test functions, and a suite of functions
from a benchmarking library. Optimization of a 5 dimensional quadratic
function (Fig.~\ref{fig:opt2}A,B) converges to the optimum for large
$S$, and converges outside of the optimum for smaller $S$. Selection
strength $t$ increases exponentially with the number of evaluations
when the algorithm is converging. On a non-convex 2 dimensional Rosenbrock
function (Fig.~\ref{fig:opt2}C,D,E) selection is tuned according
to the curvature of the landscape over the course of the optimization.
Initially, selection increases rapidly as the population falls into
the main valley, then selection slows down as it moves along the flatter
part of the valley floor, and finally speeds up again as it converges
to a quadratic optimum. 

To more rigorously assess performance, we tested QGA on noiseless
test functions from the blackbox optimization benchmark library \cite{hansen_real-parameter_2009}
(implemented in \cite{nikohansen_cma-es/pycma_2019}), which implement
an ensemble of 'instances' of each function with randomized locations
of optima. On 5 dimensional quadratic and Rosenbrock functions, the
chance of convergence to the optimum, and the average number of function
evaluations is sensitive to the hyper-parameter $S$, with low $S$
leading to premature convergence, and high $S$ leading to a high
number of function evaluations (Fig.~\ref{fig:benchmark1} red lines).
Performance is significantly improved with the modified recombination
operator which replaces $\boldsymbol{\mu}_{t}$ in Eq.~\ref{eq:recomb}
with the best variant seen thus far $\boldsymbol{\mu}_{\infty}$.
This modified algorithm is able to make larger steps towards optimum
with lower values of $S$ without prematurely converging (Fig.~S\ref{fig:benchmark1}
blue lines). 

\begin{figure*}
\begin{centering}
\includegraphics[width=0.75\textwidth]{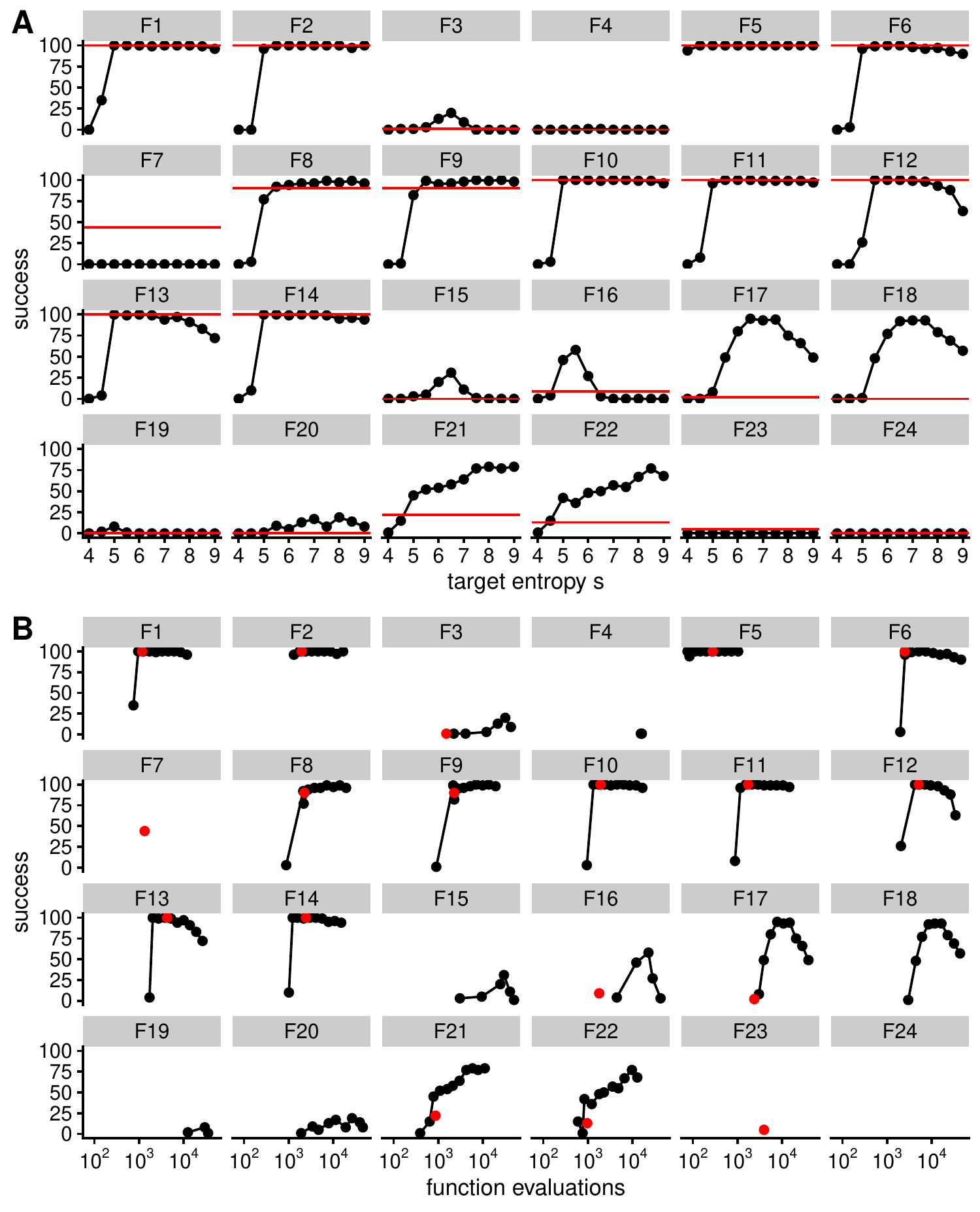}
\par\end{centering}
\caption{\label{fig:benchmarkall}QGA performance is similar to CMA-ES on 24
noiseless functions from the blackbox optimization benchmark library
with dimension 5 and different values of target diversity $S$, with
QGA (black), and CMA-ES (red). The best choice of $S$ maximizes the
chance of convergence and minimizes the number of function evaluations.
A) Successful convergence as a function of $S$ in 100 instances with
randomized optima, where success is when the lowest value found is
within $10^{-8}$ of the true minimum after less than $5\times10^{4}$
evaluations. B) Success as function of the median number of function
evaluations conditional on success, with $S$ as in A. Both algorithms
had initial conditions of mean zero and three standard deviations
in each dimension. F9 and F10 are the familiar Rosenbrock and ellipsoid
functions respectively.}
\end{figure*}

We tested QGA on all 24 noiseless test functions with the modified
recombination operator over many values of $S$ and compare to CMA-ES
\cite{nikohansen_cma-es/pycma_2019} (Fig.~\ref{fig:benchmarkall}).
Overall the performance of QGA with a good choice of $S$ is very
similar to CMA-ES, which may be expected due to their fundamental
similarities. Both algorithms perform well on functions with strong
global structure, where quadratic approximations may work well, and
perform worse on functions with highly multimodal structure. QGA has
higher chances of convergence for some of the multimodal functions
(F15-F22). For the step ellipsoidal function (F7), QGA fails completely
due to our test for premature convergence (see Methods), although
CMA-ES also performs poorly. 

For these benchmarks, QGA requires more fine tuning of its hyper-parameter
than CMA-ES. However, QGA is simpler conceptually and in implementation
compared to CMA-ES and NES. Those algorithms are somewhat complex
and involve matrix inversions or decompositions, as well as adaptive
learning rate schemes. QGA does not require any linear algebra and
does not store a covariance matrix explicitly, which may make it possible
to use on higher dimensional problems, where storage of a covariance
matrix may be an issue. Additionally, QGA naturally incorporates information
from the history of the optimization, whereas CMA-ES/NES has ``generations''
of samples and incorporates past information more heuristically. 

In comparison to GA, QGA has deterministic selection which is much
more efficient than stochastic fitness proportional selection. In
addition, the recombination operator preserves relevant correlations
between dimensions, in contrast to typical crossover operators. 

We have shown that selection steps in ES, GA, and population genetics
are intimately related, and developed a way to control the increase
in selection. Our method of increasing selection by fixing the population
entropy is simple and adaptive, yet its implications are not entirely
clear, and there may be other strategies that tune selection more
efficiently. The challenge is to find observables that indicate premature
convergence early enough to be able to continually adapt the diversity
of the population. 

Our recombination operator assumes an underlying normal distribution
to be effective. Following our scheme, a more general model-based
optimization with natural gradient updates would iterate between tuning
selection with exponential weights, and training some chosen generative
model with that weighted data. The benefit of using natural selection
is that the selection step does not require any knowledge of the generative
model, including derivatives with respect to its parameters. Such
an algorithm may be useful in generating high fitness variants for
difficult problems, including complicated discrete spaces such as
in the directed evolution of proteins.

\subsection*{Methods}

\paragraph*{Algorithm details}

Inputs are the chosen target entropy $S$ (in base 2), and the mean
and variance of the initial normal distribution from which $K_{0}=2^{S+1}$
variants are drawn. In the selection step, we calculate $p_{i}(t)=e^{tf(\boldsymbol{x}_{i})}/Z_{t},$
where $Z_{t}=\sum e^{tf(\boldsymbol{x}_{i})}$ and $t$ is incremented
or decremented geometrically by a small value until the entropy matches
$S$. In the recombination step, a new variant is generated with an
unbiased version of eq.~\ref{eq:recomb}, where $p_{i}(t)\to\frac{p_{i}(t)}{1-\sum_{i}e^{2tf(\boldsymbol{x}_{i})}}$.
To simplify the implementation, the list of variants is held at a
fixed size $K_{0}$ by replacing the variant with the lowest probability
with the new recombinant. Selection and recombination are iterated
until the desired convergence criteria are met, or the following test
for premature convergence passes. Detecting premature convergence
is challenging, and the adequate solution we found is based on comparing
fitness values. If the relative fitness differences are comparable
to the machine error, the result of recombination is that the same
fitness value may appear more than once for different values of $\boldsymbol{x}$.
Therefore, the algorithm terminates if duplicate fitness values are
found in the population. Code available at \href{https://github.com/jotwin/qga}{github.com/jotwin/qga}

\paragraph*{Test functions }

The test ellipsoid is $f(x)=(x_{1}+2x_{2}+3x_{3}+4x_{4}+5x_{5})^{2}$,
and populations were initialized with 200 random variants drawn from
a normal distribution with mean and variance equal to all ones, so
that populations are not centered on the optimum at zero. The test
Rosenbrock function is $f(x)=(1-x_{1})^{2}+100(x_{2}-x_{1}^{2})^{2}$,
and populations were initialized with 200 random variants drawn from
a normal distribution with mean (0,1) and standard deviation (0.25,
0.25). 


\begin{thebibliography}{10}

\bibitem{simon_evolutionary_2013}
Dan Simon.
\newblock {\em Evolutionary {{Optimization Algorithms}}}.
\newblock {John Wiley \& Sons}, June 2013.

\bibitem{gillespie_population_2004}
John~H. Gillespie.
\newblock {\em Population {{Genetics}}: {{A Concise Guide}}}.
\newblock {JHU Press}, 2004.

\bibitem{hauschild_introduction_2011}
Mark Hauschild and Martin Pelikan.
\newblock An introduction and survey of estimation of distribution algorithms.
\newblock {\em Swarm and Evolutionary Computation}, 1(3):111--128, September
  2011.

\bibitem{amari_why_1998}
S.~Amari and S.~C. Douglas.
\newblock Why natural gradient?
\newblock In {\em Proceedings of the 1998 {{IEEE International Conference}} on
  {{Acoustics}}, {{Speech}} and {{Signal Processing}}, 1998}, volume~2, pages
  1213--1216 vol.2, May 1998.

\bibitem{hansen_adapting_1996}
N.~Hansen and A.~Ostermeier.
\newblock Adapting arbitrary normal mutation distributions in evolution
  strategies: The covariance matrix adaptation.
\newblock In {\em Proceedings of {{IEEE International Conference}} on
  {{Evolutionary Computation}}}, pages 312--317, May 1996.

\bibitem{hansen_cma_2006}
Nikolaus Hansen.
\newblock The {{CMA Evolution Strategy}}: {{A Comparing Review}}.
\newblock In Jose~A. Lozano, Pedro Larra{\~n}aga, I{\~n}aki Inza, and Endika
  Bengoetxea, editors, {\em Towards a {{New Evolutionary Computation}}:
  {{Advances}} in the {{Estimation}} of {{Distribution Algorithms}}}, Studies
  in {{Fuzziness}} and {{Soft Computing}}, pages 75--102. {Springer}, {Berlin,
  Heidelberg}, 2006.

\bibitem{wierstra_natural_2008}
Daan Wierstra, Tom Schaul, Jan Peters, and Juergen Schmidhuber.
\newblock Natural {{Evolution Strategies}}.
\newblock In {\em 2008 {{IEEE Congress}} on {{Evolutionary Computation}}
  ({{IEEE World Congress}} on {{Computational Intelligence}})}, pages
  3381--3387, June 2008.

\bibitem{sun_efficient_2009}
Yi~Sun, Daan Wierstra, Tom Schaul, and Juergen Schmidhuber.
\newblock Efficient {{Natural Evolution Strategies}}.
\newblock In {\em Proceedings of the 11th {{Annual Conference}} on {{Genetic}}
  and {{Evolutionary Computation}}}, {{GECCO}} '09, pages 539--546, {New York,
  NY, USA}, 2009. {ACM}.

\bibitem{glasmachers_exponential_2010}
Tobias Glasmachers, Tom Schaul, Sun Yi, Daan Wierstra, and J{\"u}rgen
  Schmidhuber.
\newblock Exponential {{Natural Evolution Strategies}}.
\newblock In {\em Proceedings of the 12th {{Annual Conference}} on {{Genetic}}
  and {{Evolutionary Computation}}}, {{GECCO}} '10, pages 393--400, {New York,
  NY, USA}, 2010. {ACM}.

\bibitem{wierstra_natural_2014}
Daan Wierstra, Tom Schaul, Tobias Glasmachers, Yi~Sun, Jan Peters, and
  J{\"u}rgen Schmidhuber.
\newblock Natural {{Evolution Strategies}}.
\newblock {\em J. Mach. Learn. Res.}, 15(1):949--980, January 2014.

\bibitem{ollivier_information-geometric_2011}
Yann Ollivier, Ludovic Arnold, Anne Auger, and Nikolaus Hansen.
\newblock Information-{{Geometric Optimization Algorithms}}: {{A Unifying
  Picture}} via {{Invariance Principles}}.
\newblock {\em arXiv:1106.3708 [math]}, June 2011.

\bibitem{amari_information_2016}
Shun-ichi Amari.
\newblock {\em Information {{Geometry}} and {{Its Applications}}}, volume 194
  of {\em Applied {{Mathematical Sciences}}}.
\newblock {Springer Japan}, {Tokyo}, 2016.

\bibitem{harper_information_2009}
Marc Harper.
\newblock Information {{Geometry}} and {{Evolutionary Game Theory}}.
\newblock {\em arXiv:0911.1383 [cs, math, nlin]}, November 2009.

\bibitem{mustonen_fitness_2010}
Ville Mustonen and Michael L{\"a}ssig.
\newblock Fitness flux and ubiquity of adaptive evolution.
\newblock {\em Proc Natl Acad Sci U S A}, 107(9):4248--4253, March 2010.

\bibitem{lande_quantitative_1979}
Russell Lande.
\newblock Quantitative {{Genetic Analysis}} of {{Multivariate Evolution}},
  {{Applied}} to {{Brain}}: {{Body Size Allometry}}.
\newblock {\em Evolution}, 33(1):402--416, 1979.

\bibitem{nourmohammad_evolution_2013}
Armita Nourmohammad, Stephan Schiffels, and Michael L{\"a}ssig.
\newblock Evolution of molecular phenotypes under stabilizing selection.
\newblock {\em J. Stat. Mech.}, 2013(01):P01012, 2013.

\bibitem{kirkpatrick_optimization_1983}
S.~Kirkpatrick, C.~D. Gelatt, and M.~P. Vecchi.
\newblock Optimization by {{Simulated Annealing}}.
\newblock {\em Science}, 220(4598):671--680, May 1983.

\bibitem{amey_analysis_2018}
Christopher Amey and Jonathan Machta.
\newblock Analysis and optimization of population annealing.
\newblock {\em Phys. Rev. E}, 97(3):033301, March 2018.

\bibitem{lande_genetic_1980}
Russell Lande.
\newblock The {{Genetic Covariance Between Characters Maintained}} by
  {{Pleiotropic Mutations}}.
\newblock {\em Genetics}, 94(1):203--215, January 1980.

\bibitem{neher_statistical_2011}
Richard~A. Neher and Boris~I. Shraiman.
\newblock Statistical genetics and evolution of quantitative traits.
\newblock {\em Reviews of Modern Physics}, 83(4):1283--1300, November 2011.

\bibitem{storn_differential_1997}
Rainer Storn and Kenneth Price.
\newblock Differential {{Evolution}} \textendash{} {{A Simple}} and {{Efficient
  Heuristic}} for global {{Optimization}} over {{Continuous Spaces}}.
\newblock {\em Journal of Global Optimization}, 11(4):341--359, December 1997.

\bibitem{hansen_real-parameter_2009}
Nikolaus Hansen, Anne Auger, Steffen Finck, and Raymond Ros.
\newblock Real-{{Parameter Black}}-{{Box Optimization Benchmarking}} 2009:
  {{Experimental Setup}}.
\newblock Report, 2009.

\bibitem{nikohansen_cma-es/pycma_2019}
{nikohansen}, Youhei Akimoto, Dimo Brockhoff, and Matthew Chan.
\newblock {{CMA}}-{{ES}}/pycma: R2.7.0.
\newblock Zenodo, April 2019.

\end{thebibliography}

\clearpage{}

\appendix
\onecolumngrid

\section{Natural selection gradient \label{sec:Selection}}

We begin by following \cite{mustonen_fitness_2010}. A population
has $i=1...K$ variants with frequencies $p_{i}$ that evolve according
to
\[
\frac{dp_{i}}{dt}=p_{i}(f(\boldsymbol{x}_{i})-F),
\]
with fitness $f$, phenotype $\boldsymbol{x}_{i}$, and mean fitness
$F=\sum_{i=1}^{K}p_{i}f(x_{i})$. There are $K-1$ independent variables,
so we choose the $K$th variant as the reference and

\[
p_{K}=1-\sum_{i=1}^{K-1}p_{i}.
\]
Eliminating this variable results in
\[
\frac{dp_{i}}{dt}=\sum_{j=1}^{K-1}g_{ij}^{-1}\frac{\partial F}{\partial p_{j}},
\]
 with 
\[
g_{ij}^{-1}=p_{i}\delta_{ij}-p_{i}p_{j}
\]
where $\delta_{ij}$ is the kronecker delta. $g_{ij}^{-1}$ is also
the covariance of a categorical distribution, and appears in the generalized
Kimura diffusion equation. Note that indices are over the $K-1$ variables.

The inverse of this matrix is the Fisher information metric 
\begin{align*}
g_{ij} & =\sum_{k=1}^{K}p_{k}\left(\frac{\partial}{\partial p_{i}}\log p_{k}\right)\left(\frac{\partial}{\partial p_{j}}\log p_{k}\right)\\
 & =\sum_{k=1}^{K}\frac{1}{p_{k}}\frac{\partial p_{k}}{\partial p_{i}}\frac{\partial p_{k}}{\partial p_{j}}\\
 & =\frac{1}{p_{K}}+\frac{1}{p_{i}}\delta_{ij}
\end{align*}

If $p_{i}$ describe a normally distributed phenotype $\boldsymbol{x}$,
we can project onto the phenotype mean and covariance
\[
\boldsymbol{\mu}=\sum_{i=1}^{K-1}p_{i}(\boldsymbol{x}_{i}-\boldsymbol{x}_{k})+\boldsymbol{x}_{k}
\]
\[
\boldsymbol{\Sigma}=\sum_{i=1}^{K-1}p_{i}\left[(\boldsymbol{x}_{i}-\boldsymbol{\mu})^{2}-(\boldsymbol{x}_{k}-\boldsymbol{\mu})^{2}\right]+(\boldsymbol{x}_{k}-\boldsymbol{\mu})^{2}
\]
 with multivariate phenotypes $\boldsymbol{x}_{i}$ as vectors of
dimension $D$, and shorthand for the matrix $\boldsymbol{b}^{2}=\boldsymbol{bb}^{\top}$
given a vector $\boldsymbol{b}$. The derivatives of mean fitness
can be decomposed by the chain rule
\begin{align*}
\frac{\partial F}{\partial p_{j}} & =\nabla_{\boldsymbol{\mu}}F\cdot\frac{\partial\boldsymbol{\mu}}{\partial p_{j}}+\nabla_{\boldsymbol{\Sigma}}F\cdot\frac{\partial\boldsymbol{\Sigma}}{\partial p_{j}}\\
\frac{\partial\boldsymbol{\mu}}{\partial p_{j}} & =\boldsymbol{x}_{j}-\boldsymbol{x}_{K}\\
\frac{\partial\boldsymbol{\Sigma}}{\partial p_{j}} & =(\boldsymbol{x}_{j}-\boldsymbol{\mu})^{2}-(\boldsymbol{x}_{k}-\boldsymbol{\mu})^{2}
\end{align*}
where $\nabla_{\boldsymbol{b}}$ is the vector of partial derivatives
with respect to parameter $\boldsymbol{b}$. Combining the above equations
results in 
\begin{align*}
\frac{d\boldsymbol{\mu}}{dt} & =\boldsymbol{\Sigma}\nabla_{\boldsymbol{\mu}}F
\end{align*}
\[
\frac{d\boldsymbol{\Sigma}}{dt}=2\boldsymbol{\Sigma}\boldsymbol{\Sigma}^{\top}\nabla_{\boldsymbol{\Sigma}}F
\]

\section{Selection and Newton's method\label{sec:newton}}

The integrated dynamics of the frequencies are

\[
p_{i}(t)=p_{i}(0)\frac{1}{Z_{t}}e^{tf(\boldsymbol{x_{i}})},
\]
with normalization $Z_{t}=\sum_{i}p_{i}(0)e^{tf(\boldsymbol{x}_{i})}$
The change in mean phenotype is proportional to the covariance between
phenotype and fitness, (Price's theorem) 
\begin{align*}
\boldsymbol{\mu}_{t}-\boldsymbol{\mu}_{0} & =\frac{1}{Z_{t}}\sum_{i=1}^{K}p_{i}(0)\left(\boldsymbol{x}_{i}e^{tf(\boldsymbol{x}_{i})}-\boldsymbol{\mu}_{0}Z_{t}\right).
\end{align*}

A taylor expansion of $f(\boldsymbol{x})$ around $\boldsymbol{\mu}_{0}$
to second order is
\[
f(\boldsymbol{x})\approx f(\boldsymbol{\mu}_{0})+(\boldsymbol{x}-\boldsymbol{\mu}_{0})\nabla f-\frac{1}{2}(\boldsymbol{x}-\boldsymbol{\mu}_{0})^{2}\cdot\boldsymbol{C},
\]
where $\nabla f$ is the vector of partial derivates of fitness evaluated
at $\boldsymbol{\mu}_{0}$, and $\boldsymbol{C}$ is the curvature,
that is the negative of the matrix of second partial derivatives evaluated
at $\boldsymbol{\mu}_{0}$, and we use a shorthand for the outer product,
e.g, $\boldsymbol{b}^{2}=\boldsymbol{b}\boldsymbol{b}^{\top}$, and
$\boldsymbol{b}^{2}\cdot A=\boldsymbol{b}^{\top}A\boldsymbol{b}$,
for an arbitrary matrix $A$. The expansion is more conveniently written
as 
\[
f(x)\approx F^{*}-\frac{1}{2}(\boldsymbol{x}-\boldsymbol{x}^{*})^{2}\cdot\boldsymbol{C}
\]
where $F^{*}$ and $\boldsymbol{x}^{*}$ are the approximate optimum
fitness and phenotype respectively, and $\boldsymbol{C}(\boldsymbol{x}^{*}-\boldsymbol{\mu}_{0})=\nabla f$.

If phenotypes are initially normally distributed then they remain
normal after selection with the quadratic approximation of the fitness
landscape, as can be seen from the integrated replicator dynamics.
The resulting mean and inverse covariance are
\begin{align*}
\boldsymbol{\mu}_{t} & =\boldsymbol{\Sigma}_{t}\left(\boldsymbol{\Sigma}_{0}^{-1}\mu_{0}+t\boldsymbol{C}\boldsymbol{x}^{*}\right)\\
\boldsymbol{\Sigma}_{t}^{-1} & =\boldsymbol{\Sigma}_{0}^{-1}+t\boldsymbol{C}
\end{align*}
The change in the mean is 
\begin{align*}
\boldsymbol{\mu}_{t}-\boldsymbol{\mu}_{0} & =\boldsymbol{\Sigma}_{t}t\nabla f\\
 & =\left(\frac{1}{t}\boldsymbol{\Sigma}_{0}^{-1}+\boldsymbol{C}\right)^{-1}\nabla f.
\end{align*}
Under strong selection in a quadratic fitness landscape the covariance
shrinks to zero at a rate
\[
\boldsymbol{\Sigma}_{t}\to\frac{\boldsymbol{C}^{-1}}{t}
\]

\section{Genetic load\label{sec:Genetic-load}}

Fitness is distributed as a noncentral chi-squared distribution, since
fitness is a sum of squared values that are normally distributed.
The difference between the optimum and the mean fitness:
\begin{align}
F^{*}-F & =\frac{1}{2}\sum_{i}p_{i}(\boldsymbol{x}_{i}-\boldsymbol{x}^{*})^{2}\cdot\boldsymbol{C}\nonumber \\
 & =\frac{1}{2}\left(\boldsymbol{\Sigma}_{t}+(\boldsymbol{\mu}_{t}-\boldsymbol{x}^{*})^{2}\right)\cdot\boldsymbol{C}\label{eq:load}
\end{align}
which decomposes into an effective degrees of freedom $Tr(\boldsymbol{\Sigma}_{t}\boldsymbol{C})$,
and a noncentrality parameter $(\boldsymbol{\mu}_{t}-\boldsymbol{x}^{*})^{2}\cdot\boldsymbol{C}$.
For strong selection this fitness difference aproaches zero as 
\[
F^{*}-F\to\frac{D}{2t}+\mathcal{O}(t^{-2})
\]
 where $D$ is the dimension of $x$.

\section{Mutations\label{sec:Mutations}}

If mutations are normally distributed around each focal phenotype,
the total covariance of all mutants will be the phenotype covariance
plus the mutational covariance $\Sigma+\Sigma_{m}$. The associated
mutational load is $\textrm{Tr}(\boldsymbol{\Sigma}_{m}\boldsymbol{C})$,
and if mutations are standard normal, then the mutational load is
the sum of the eigenvalues of $\boldsymbol{C}$, or roughly the number
of dimensions where fitness is sharply curved and not flat.

\section{Gradient step in SSWM moran model\label{sec:SSWM}}

In strong selection weak mutation dynamics {[}{]}, a population is
monomorphic (only one type), and changes type $\boldsymbol{x}$ by
mutation-fixation events. A mutant has a fitness difference $f(\boldsymbol{x}^{\prime})-f(\boldsymbol{x})\approx\Delta\boldsymbol{x}\nabla f\equiv s$,
and the probability of fixation is 
\[
P_{\text{fix}}(\boldsymbol{x}^{\prime}|\boldsymbol{x})=\begin{cases}
2s & s>0\\
0 & s\le0
\end{cases}.
\]
The expected change in $\boldsymbol{x}$ is
\[
E(\Delta x)=\int\Delta\boldsymbol{x}P_{\text{fix}}(\boldsymbol{x}^{\prime}|\boldsymbol{x})P_{m}(\boldsymbol{x}^{\prime}|\boldsymbol{x})d\boldsymbol{x}^{\prime},
\]
where $P_{m}(x^{\prime}|x)$ is the mutation distribution around $x$.
For a normal mutation distribution $P_{m}(\boldsymbol{x}^{\prime}|\boldsymbol{x})=\mathcal{N}(\boldsymbol{x},\Sigma_{m})$,
\[
E(\Delta x)=\Sigma_{m}\nabla f
\]

\clearpage{}

\setcounter{figure}{0}
\renewcommand{\thefigure}{S\arabic{figure}}
\begin{figure*}
\begin{centering}
\includegraphics[width=1\textwidth]{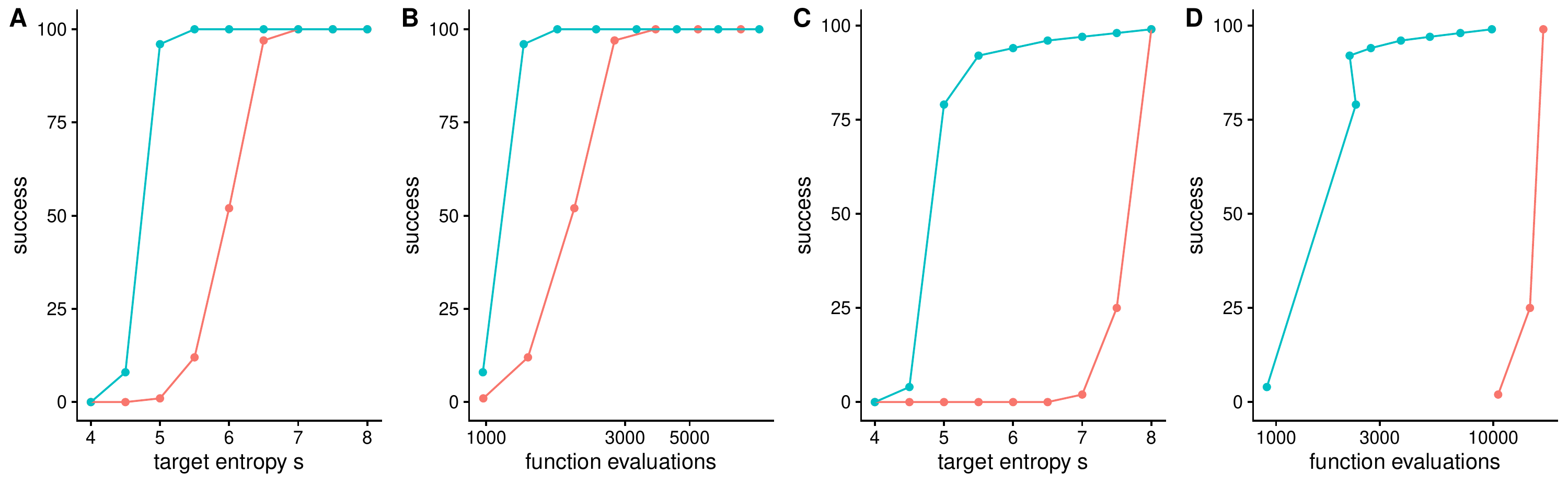}
\par\end{centering}
\caption{\label{fig:benchmark1}QGA performance on 5 dimensional quadratic
(A,B) and Rosenbrock (C,D) functions from the blackbox optimization
benchmark library with different values of target diversity $S$.
The best choice of $S$ maximizes the chance of convergence and minimizes
the number of function evaluations. A,C) Successful convergence as
a function of $S$ in 100 instances with randomized optima, where
success is when the lowest value found is within $10^{-8}$ of the
true minimum after less than $5\times10^{4}$ evaluations. B,D) Success
as function of the median number of function evaluations conditional
on success, with $S$ as in A,C. Red lines indicate standard recombination,
Eq.~\ref{eq:recomb}, and blue lines indicate modified recombination
(see text). Each population was initialized with mean zero and standard
deviation of 3 in each dimension.}
\end{figure*}

\end{document}